\documentclass[aps,prb,twocolumn,showpacs,floatfix,a4paper]{revtex4}
\usepackage{graphicx}
\usepackage[intlimits]{amsmath}
\usepackage{color}

\begin{document}

\title{Large scale behavior of
  the energy spectra of the quantum random antiferromagnetic Ising chain
  with mixed transverse and longitudinal fields}
\author{Richard Berkovits}
\affiliation{Department of Physics, Jack and Pearl Resnick Institute, Bar-Ilan University, Ramat-Gan 52900, Israel}

\begin{abstract}
  In recent years it became  clear that the metallic regime of systems that
  exhibit a many body localization (MBL) behavior show properties which are
  quite different than the vanilla metallic region of
  the single particle Anderson regime. Here we show that the large scale
  energy spectrum of a canonical microscopical model featuring MBL,
  displays a non-universal behavior at intermediate scales, which is distinct
  from  the deviation from universality seen in the single particle
  Anderson regime. The crucial  step in revealing  this behavior is a global
  unfolding of the spectrum performed using the singular value decomposition
  (SVD) which takes into account the sample to sample fluctuations of the
  spectra. The spectrum properties may be observed directly in the
  singular value amplitudes via the scree plot, or by using the SVD to
  unfold the spectra and then perform a number of states variance calculation.
  Both methods reveal an intermediate scale of energies which follow
  super Posissonian statistics.
\end{abstract}

%\pacs{73.21.Hb,71.15.Rn,71.10.Hf,71.27.+a}

\maketitle

\section{Introduction}

Many-body localization (MBL) \cite{r1,r2} has captured the imagination
of researchers since its inception more than a decade and a half ago.
Once interactions are introduced to a many-particle system for which
all single particle states are localized, a parameter region where
the many-particle states are extended should appear, as expected
from the many-body thermalization hypothesis \cite{r3,r4}. Other
regions of the parameter space remain localized even in the
presence of interactions. Almost immediately, an effort to
identify the transition point by analyzing the spectra of microscopic
models of disordered interacting many-particle systems began.
The spectra of 1D spin chains and electronic models
\cite{oganesyan07,pal10,luitz15,modaini15,chanda20}
were probed in order to identify a signature of a
transition (or crossover) between the two regions.
Nevertheless, despite much effort a definitive answer remains elusive.

For microscopic models of MBL one runs into
an insurmountable obstacle in analyzing the energy spectra. The Hilbert
space grows exponentially and for conventional computers it is hard to imagine
that one will reach large enough systems for which the analysis of the
spectra will give an indisputable finite size scaling. Nevertheless, there
is still a point in looking into the spectral properties of small 
microscopic models, for two main reasons. The first, is that although the
systems studied are small, there are nevertheless some behaviors which emerge
in a robust form even for these sizes. Although it might not be possible to
prove that these behaviors survive in the thermodynamical limit, it
is still worthwhile to understand them \cite{monteiro21}.
Second, many current experimental
studies searching for a signature of the MBL \cite{ex1,ex2,ex3,ex4,ex5}
are performed on systems of similar small size.

Here we would like to examine a particular microscopical model
of a quantum random antiferromagnetic Ising chain with mixed transverse
and longitudinal fields, sometimes referred in the MBL literature as the
Imbrie model. The ground state of this model has been known
to exhibit a rich phase diagram \cite{fisher95,igloi05,lajko20},
and recently the model has garnered considerable interest in the context of
MBL \cite{imbrie16,imbrie16a,biroli20,tarzia20,abanin21,tomasi21}.
This interest stems from the assertion that under some assumptions,  it
is possible to rigorously show that it undergoes a
MBL transition from metallic to localized behavior as
disorder increases \cite{imbrie16a}.

What is the nature of the extended region of the Imbrie model?
Is this region analogues to the single particle Anderson metallic phase?
These are the questions we would like to address in this paper.
For single particle Anderson metallic regime, the energy spectrum
follows the random matrix predictions
\cite{mehta91,shklovskii93,ghur98,alhassid00,mirlin00,evers08} up to an energy
scale known as the Thouless energy \cite{altshuler86}, above
which a different behavior is observed, where the Thouless energy
$E_{Th}=\hbar D/L^2=g \delta$ ($D$ is the diffusion constant, $L$, is
the linear dimension, $g$ is the dimensionless conductance, and $\delta$
the average level spacing). 
The physical origin of the
Thouless energy is the time needed for a wave packet to cover the whole sample
known as the Thouless time $t_{Th}=\hbar/E_{Th}=L^2/D$. At shorter times
(larger energy scales) the system is not ergodic, hence the different
energy spectrum behavior at this scale.

The generalized Rosenzweig-Porter random matrix
model (GRP) \cite{rosenzweig60,kravtosov15}
is probably the simplest random matrix model which
shows three distinct phases: Localized at
strong disorder, non-ergodic
extended (NEE) phase for intermediate disorder, and
a fully ergodic extended phase at weak disorder.
The NEE phase exhibits unusual features such as
fractality
of the wave functions \cite{kravtosov15,monthus17,kravtsov18,bogomolny18,nosov19,pino19,detomasi19,khaymovich20}, and super Poissonian behavior of
the energy spectrum at intermediate energy scales \cite{detomasi19,berkovits20}.
Focusing on the energy spectrum, one discovers that
the nearest neighbor statistics (small energy scale, corresponding
to long times) is indistinguishable from the extended metallic phase,
while for intermediate energy scales a 
super-Poissonian
behavior of the n-th level spacing distribution has been observed
\cite{detomasi19}. Examining the singular value decomposition (SVD) of
the  spectrum of an ensemble  of realizations supports this conclusion.
Moreover, the SVD amplitude scree plot, which in the NEE phase show
three different regimes as function of the mode number (essentially
inverse energy, where low modes correspond to large energy scales).
High modes (small energy scales) show a Wigner behavior, then it crosses
to a super-Poissonian behavior for the  intermediate range of modes, finally
switching at low modes (large energy scales) to a Poisson form
\cite{berkovits20}. Thus two transition energies in the
spectrum emerge. The lower transition energy corresponds
to a transition from a universal Wigner behavior typical to the metallic 
regime, to a non-universal NEE regime.
Following previous works \cite{garcia16},
the energy at which this transition occurs will be termed the  
Thouless energy. It is important though to note that although the same
terminology for the Thouless energy as for the single particle Anderson
transition is used, it does not necessarily mean that the same
physics is behind it. This will be discussed further on.
%{\red Similar
%in a sense to the transition from  universal behavior at short energy
%scales to non-universal behavior at  larger scales occurring in
%single particle metallic
%systems \cite{altshuler86},
%and therefore term this transition energy as the Thouless energy.}
The second transition between the NEE regime and Poisson like behavior
behavior has no direct analogue in the single particle metallic systems.
One must keep in mind that for any system
the energy spectrum on large scales is determined by the global band structure.
which is captured in the first few modes in the scree plot.
The NEE behavior has an additional time scale which is the onset of the extended
behavior. Thus, a region  of small modes of the SVD will be needed to
capture the pre-extended region. The
energy for which the extended behavior is manifested is the
second transition energy and will be termed, $E_{Ex}$,
the extended energy. 
Much of the interest in the GRP model stems from the proposal that it
might capture properties relevant to MBL systems.
Indeed we will demonstrate that these two transition energies
emerge also for the Imbrie model, and the meaning of
this large energy scale will be discussed further on.

One of the most interesting questions investigated
by the MBL community
\cite{deluca14,deluca13,altshuler16,tikhonov16,pino16,pino17,torres17,biroli18,tikhonov19,tikhonov19a,faoro19,kechedzhi20,biroli20}
is the nature of the metallic regime close to the localized regime.
This region exhibits
at intermediate times (intermediate energy scales) different behavior
than expected from the canonical Anderson transition.
For example, the time evolution of the system is
sub-diffusive and relaxation toward equilibrium is anomalously slow
\cite{barlev15,agarwal15,torres15,luitz16,luitz16a,luitz17,agarwal17,bera17,dogge18}, a behavior that might have been seen also in experiments
\cite{ex1,ex6,ex7}. The same region also exhibits fractal behavior of the
eigenfunctions \cite{tomasi21}.

The fractal behavior of the eigenfunctions as well as the sub-diffusive
time evolution have several complementary explanations. One of the 
main routes to an explanation of the NEE behavior is via the picture
localization in the Fock space, which has originally motivated the study of
MBL \cite{altshuler97,r2}. Essentially the coupling of states in the
Fock space creates a quantum random graph which leads to non ergodic behavior
and fractal structure of the states in Fock space.
This behavior has also been associated to rare regions in
the 1D systems known as the Griffiths regions \cite{agarwal15,agarwal17,gopa15,vosk15,potter15,zhang16,gopa16,berkovits18},
which could explain the sub-diffusive behavior. Nevertheless, this can
not be the whole picture since sub-diffusive behavior is seen also in systems
of higher dimentionality, both numerically \cite{barlev16,doggen20} and
experimentally \cite{ex7},
while the Griffiths regions influence is limited to 1D
systems. A complementary view suggests that the MBL transition is a
Kosterlitz  Thouless (KT) transition
\cite{thiery17,goremykina19,dumitrescu19,morningstar19,morningstar20,tomasi21}.
In this picture, 
rare regions of
extended states may appear which will lead to an avalanche delocalizing
the whole sample if disorder is not too strong.
This will lead naturally to fractal structure of the Hilbert space and to
a NEE behavior. 

Here we would
explore whether there  are signatures of super Poissonic statistics
at intermediate
energies in the Imbrie model, similar to the behavior seen in GRP.
The usual way of examining the behavior
at large energy scales is the variance of the number of levels as function
of the size  of an energy window after a local unfolding
of the energy spectrum, i.e.,
$\langle \delta^2 n(E) \rangle=\langle (n(E) - \langle n(E) \rangle)^2 \rangle$,
where $\langle \ldots \rangle$ denotes an average over an ensemble of
different realizations of disorder and $n(E)$ is the number of levels within
an energy window $E$. In the Poisson regime 
$\langle \delta^2 n(E) \rangle \sim
\langle n(E) \rangle$, while in the Wigner regime it grows logarithmic.
Deviation from the logarithmic
behavior to a stronger than linear behavior at large energies have been
seen in metallic system beyond the Thouless energy \cite{braun95,cuevas97},
the Sachdev-Ye-Kitaev (SYK) model \cite{garcia16,garcia18}, and
many body localization systems \cite{bertrand16,corps20,wang21}.
As we have shown in Refs. \onlinecite{berkovits20,berkovits21}
there are some problems in the application of the local
unfolding in systems where the local density shows strong sample
to sample fluctuations or a non-smooth band structure which may 
skew the results. In order to circumvent these problems we will use 
a different method to study the properties of
the spectra, known as singular value decomposition (SVD). This method
has been successfully applied to analyze the transition from
Wigner to Poisson statistics  in the Anderson transition
\cite{fossion13,torresv17,torresv18}, to characterizing the NEE in the
GRP model \cite{berkovits20}, to study the large energy scale spectrum
behavior beyond the Thouless  energy in metallic systems \cite{berkovits21},
and very recently to the MBL transition in the Heisenberg chain \cite{rao21}.

%^^^^^^^^^^^^^^^^^^^^^^^^^^^^^^^^^^^^^^^^^^^^^^^^^^^^^^^^^^^^^^^^

As will be discussed in detail in the appendix, SVD essentially returns
a set of modes which can be used to construct the energy spectra of the
different realizations in the ensemble. Arranging the modes according 
to the size of their amplitude squared, $\lambda_k$
(where $k=1$ is the largest), the first few $\lambda_k$
($O(1)$) correspond to global features of the spectra
\cite{fossion13,torresv17,torresv18,berkovits20}.
Thus one can globally unfold the spectra by 
filtering out these modes when reconstructing the spectrum. Then the
unfolded  spectrum  can be used to obtain  the  number variance.
A different way to obtain  a comprehensive picture of the
behavior of the energy spectrum is to plot $\lambda_k$ vs. $k$,  also known as
a scree plot. Usually, a power law behavior $\lambda \sim k^{-\alpha}$, is
detected for certain ranges of $k$. The power law exponent corresponds to the
statistics of the energy spectrum with $\alpha=2$
for the Poisson behavior, $\alpha=1$ for the Wigner regime
\cite{fossion13,torresv17,torresv18,berkovits20}.
For energies larger than $E_{Th}$ (small values of $k$)
in the metallic regime of a single particle
Anderson model  $\alpha=1+d/2$  (where $d$ is the dimensionality)
\cite{berkovits21}.
In Ref. \onlinecite{berkovits20} 
we have shown that for GRP model in the NEE phase
shows for intermediate values of $k$ a power-law behavior with $\alpha>2$.
This behavior is consistent with super-Poissonian statistics.
At large values of $k$ (short energy scales)
the singular value curve returns to the $\alpha=2$ exponent, i.e., Poisson
statistics. Thus the super Poisson behavior starts at the Thouless energy, and
terminates at $E_{Ex}$.
Moreover, as has been discussed in Ref. \onlinecite{berkovits21},
the power law of the SVD amplitude scree plot is connected to the power law
behavior of the number variance, $\langle \delta^2 n(E) \rangle \sim
\langle n(E) \rangle^{\beta}$, with $\beta=\alpha-1$, where
in the Poisson regime $\alpha=2$ and $\beta=1$, while in the Wigner
regime  $\alpha=1$ and $\beta=0$ (actually logarithmic). In the
NEE regime one expects a super Poissonian behavior of the number variance
$\beta>1$  and therefore $\alpha>2$.

In this paper we shall introduce the random antiferromagnetic Ising chain
(Imbrie model) in Sec. \ref{s2}. The dependence of the density of states
and ratio statistics (nearest neighbor level statistics)
on the strength of disorder is presented in Sec. \ref{s3}.
As expected, above a  certain strength of  disorder  
finite size scaling indicates a localized regime, while
for weaker disorder an extended regime 
emerges. 
Then (Sec. \ref{s4}) the locally unfolded spectra is used to
study the level number variance. The results deviate from RMT predictions
(whether Wigner in the extended regime or Poissonian the localized)
for higher energy scales, and in the extended regime
seem to  follow a super Poissonian behavior. Nevertheless, due to the structure
of density of states as well as to strong sample to sample fluctuations, one
must question  the validity of the local unfolding. Therefore we turn to the
SVD scree plot, in order to get a
better picture of the larger energy scale behavior of
the spectrum in Sec.\ref{s5}. The scree plot suggests that
for small energies, the system follows the expectations
garnered from the  ratio statistics. Then at a particular
mode (corresponding to the Thouless energy) the power law changes to a super
Poisson value ($\alpha>2$). For large energies
a second transition is seen in the lower modes corresponding
to $E_{Ex}$, which switches
the power lat back to
$\alpha \sim 1$.
Thus, an
intermediate range of energies with super Poisonian statistics emerges for the
extended side of the MBL. For the localized regime, no super Poissonian regime
exists and the Poisson exponent transits into a smaller exponent $\alpha<2$ for
low modes. This transition mitigates as one moves deeper into  the localized
regime. In Sec. \ref{s6} SVD is used to globally unfold the spectra
and investigate the globally unfolded level number variance, resulting in values
of $\beta$ that correspond well with the $\alpha$  deduced from the
scree plot. These results as well as further ramifications are discussed
in Sec.  \ref{s7}.

%>>>>>>>>>>>>>>>>>>>>>>>>>>>>>>>>>>>>>>>>>>>>>>>>>>>>>>>>>>>>>>>

\section{Imbrie Model}
\label{s2}

The Hamiltonian for the random antiferromagnetic Ising chain of length $L$
with mixed transverse and longitudinal fields (Imbrie model) is given by
\cite{imbrie16,imbrie16a,biroli20,tarzia20,abanin21}:

\begin{eqnarray} \label{imbrie}
  \hat H=\sum_{i=1}^{L} h_i \hat S_i^z +  \sum_{i=1}^L \gamma_i \hat S_i^x + 
  \sum_i^{L-1} J_{i} \hat S_i^z\hat S_{i+1}^z,
\end{eqnarray}
where
$S_i^a$ is the spin on site $i$ in direction $\hat a$.
$h_i$ is a random magnetic field in the $\hat z$ direction
on site $i$ drawn from a
box distribution between $-W/2$ and $W/2$, and $\gamma_i=1$.
$J_{i}$ are the nearest-neighbor spin-spin antiferromagnetic interactions
which following Abanin {\it et.al.} \cite{abanin21}  are
drawn from a
box  distribution in the range $0.8$ and $1.2$. $L$ is the length
of the chain.

The corresponding to Hilbert space has size of $2^L$,  and
we calculate the eigenvalues $E_i$ using exact diagonaliztion for $M$
realizations for a given disorder. 

\section{Small Energy Scales}
\label{s3}

As a first step we would like to calculate the nearest neighbor level spacing
statistics in order to establish for which values of disorder $W$
we see extended states. Since small energy scales for the
extended regime follow the
Wigner statistics while for the localized regime they follow Poisson
statistics, a finite size scaling of a measure probing the
nearest neighbor level statistics
should reveal  in what regime the system is. As  a measure 
we shall use the ratio statistics \cite{oganesyan07}, defined as:
\begin{eqnarray} \label{ratio}
r &=& \langle \min (r_n,r_n^{-1}) \rangle,
\\ \nonumber
r_n &=& \frac {E_n-E_{n-1}}{E_{n+1}-E_{n}},
\end{eqnarray}
where $E_n$ is the n-th eigenvalue of the Hamiltonian
and $\langle \ldots \rangle$ is
an average over different realizations of disorder and a range of eigenvalues
around the middle of the energy spectrum. This measure has the advantage
of avoiding the unfolding procedure.
For the Poisson statistics $r_s= 2 \ln 2 - 1 \cong 0.3863$,
while for the GOE Wigner distribution $r_s \cong  0.5307$.
\cite{atas13}.

\begin{figure}
\includegraphics[width=8cm,height=!]{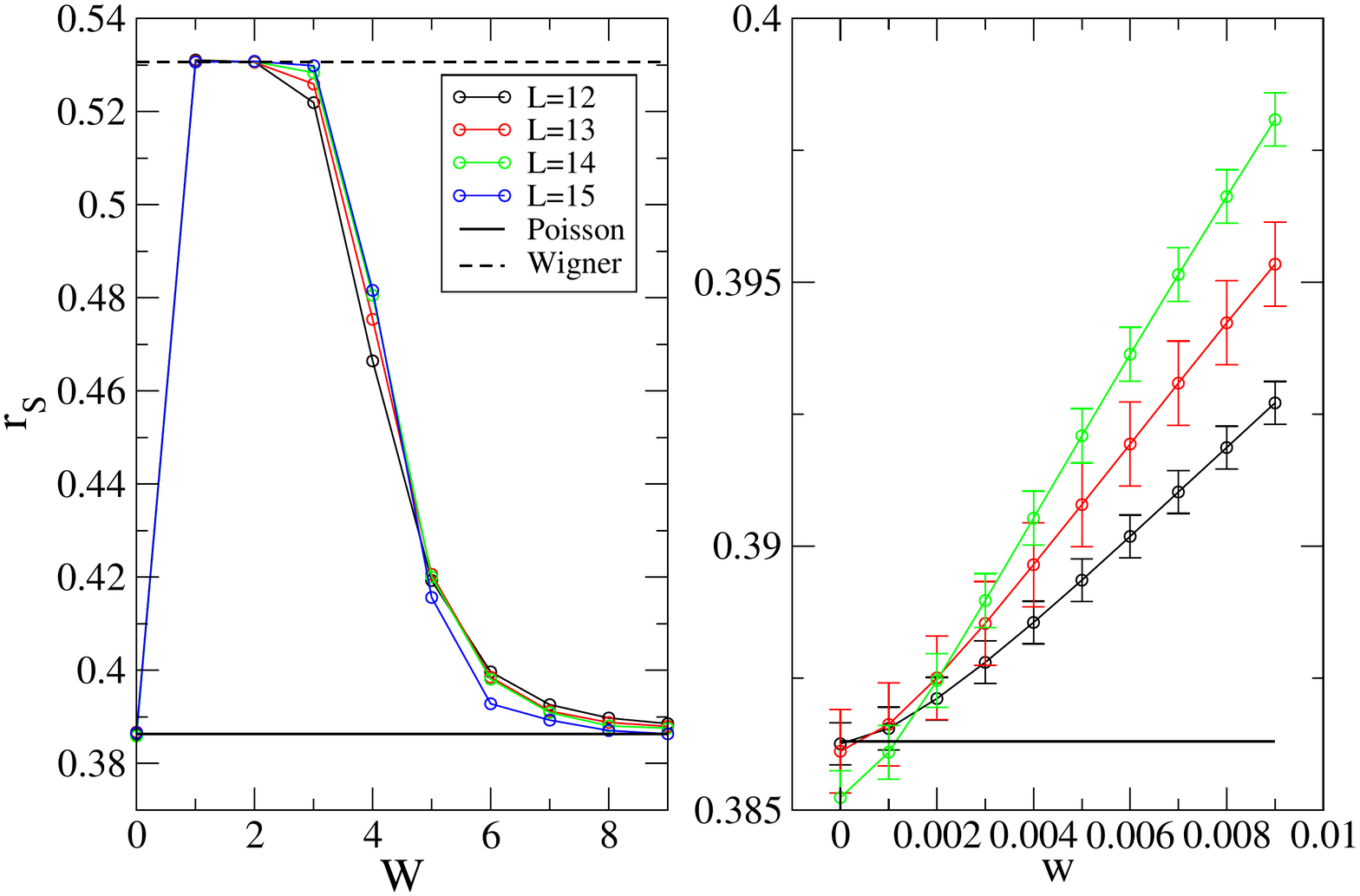}
\caption{\label{fig2}
  The ratio statistics $r_s$ (Eq. \ref{ratio})
  as function of the disorder $W$, for different system sizes $L=12,13,14,15$.
  Averaging was performed over $3000$ realizations for $L=12,13,14$,
  while for $L=15$, the averaging was performed over $1000$ realization.
  Circles correspond to the numerical results,
  while the $r_s$ values for Wigner GOE and
  Poisson  are indicated by continuous and dashed lines.
  On the left side the whole range of $W$ is presented
  while on the right side a zoom into small values of $W$ (for $L=12,13,14$)
  is shown.
}
\end{figure}

In Fig. \ref{fig2}, $r_s$,
for sample sizes, $L=12,13,14,15$
(corresponding to Hilbert space sizes of $2^L$), is presented.
The matrices were exactly diagonalized and all eigenvalues $E_n$ were
obtained. $r_s$ was
averaged over $M$ realizations, where $M=3000$ for $L=12,13,14$,
while for $L=15$, $M=1000$
realizations, using $P=2^L/2$ eigenvalues around the center of the band. 
For finite $W$ a typical transition pattern is seen:
Above $W \sim 5$
the larger is $L$ 
the lower the value of $r_s$ and the
closer to the Poisson value it becomes.
Below the value of $W \sim 5$ (except for in the vicinity of  $W=0$)
the order is the opposite,
the larger the $L$, the higher its $r_s$ value is and the closer its value is to
the GOE statistics.
All the curves seem to cross at the same value
of $W\sim 5$. Thus,
roughly speaking, the behavior of $r_s$ shows the finite size
scaling features of a
second order localization transition.
At $W=0$ the system again coalesces at the Poisson value. As
can be seen from the right panel the finite size behavior indicates
that the crossover occurs very close to $W=0$.

Since our aim in this study was to investigate
the extended regime and not to determine the nature of the transition to  the
localized regime, the values of $W$ around the intersection of the curves was
not calculated with enough points around it and the averaging was not performed
on a sufficient large number of realization to establish that the crossing
corresponds to  a  second order transition. At this point we can not be sure
that the crossing does not drift with size
or show Kosterlitz-Thouless like behavior.

\section{Number Variance with Local Unfolding}
\label{s4}

\begin{figure}
\includegraphics[width=8cm,height=!]{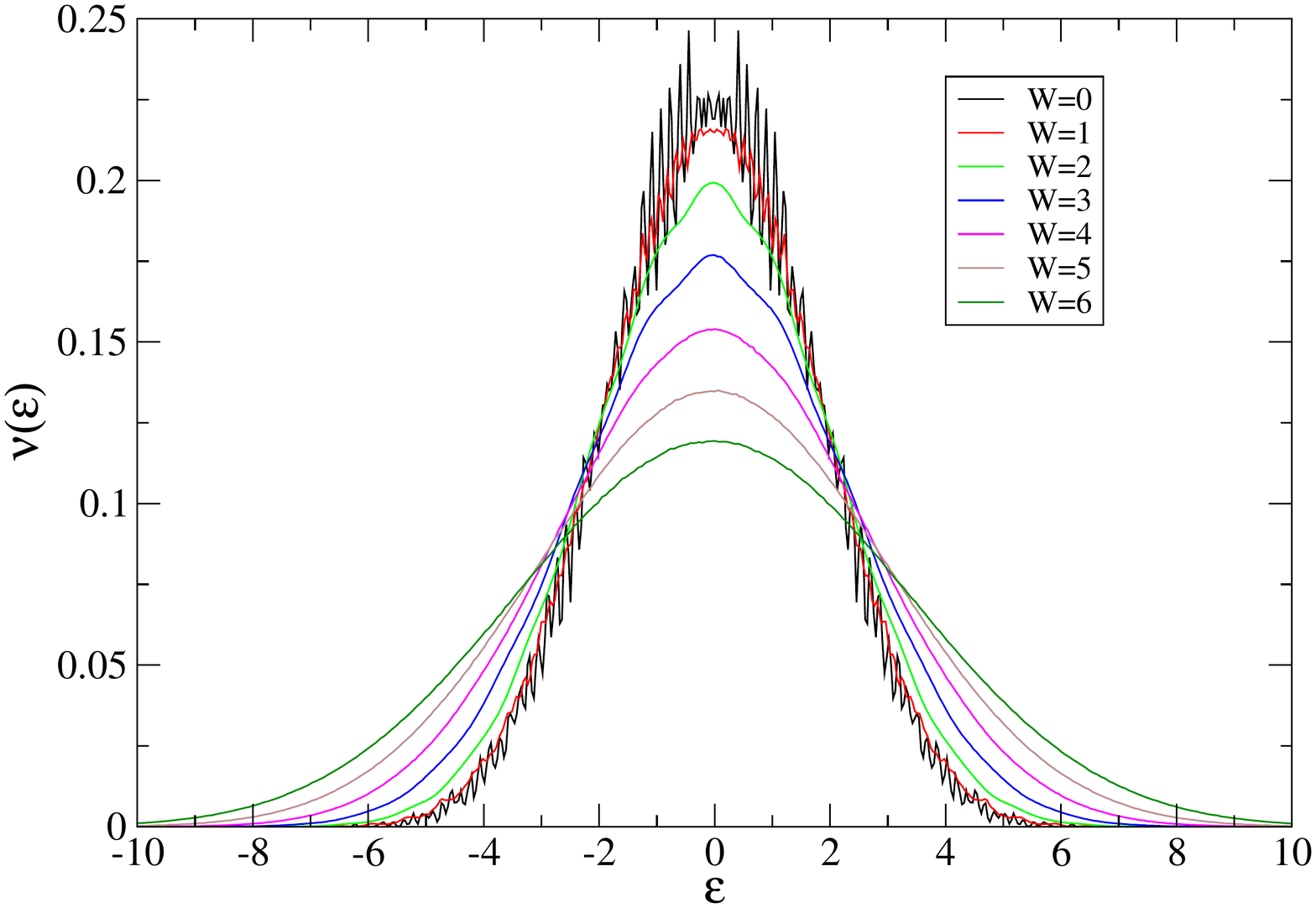}
\caption{\label{fig0}
  The density of states $\langle \nu(\varepsilon) \rangle$ as function of
  the energy $\varepsilon$
  for $L=14$ and different values of disorder averaged over
  $M=3000$ different realizations of disorder.
 }
\end{figure}

We start by plotting the average density of states
$\nu(\varepsilon)$ for different values
of disorder. As can be seen in Fig. \ref{fig0} the level density widens as
expected when the disorder increases. Moreover, it is also apparent that the
density becomes more smooth as $W$ increases. For $W<2$ some additional (quasi-)
regular structure of the density is seen.

\begin{figure}
\includegraphics[width=8cm,height=!]{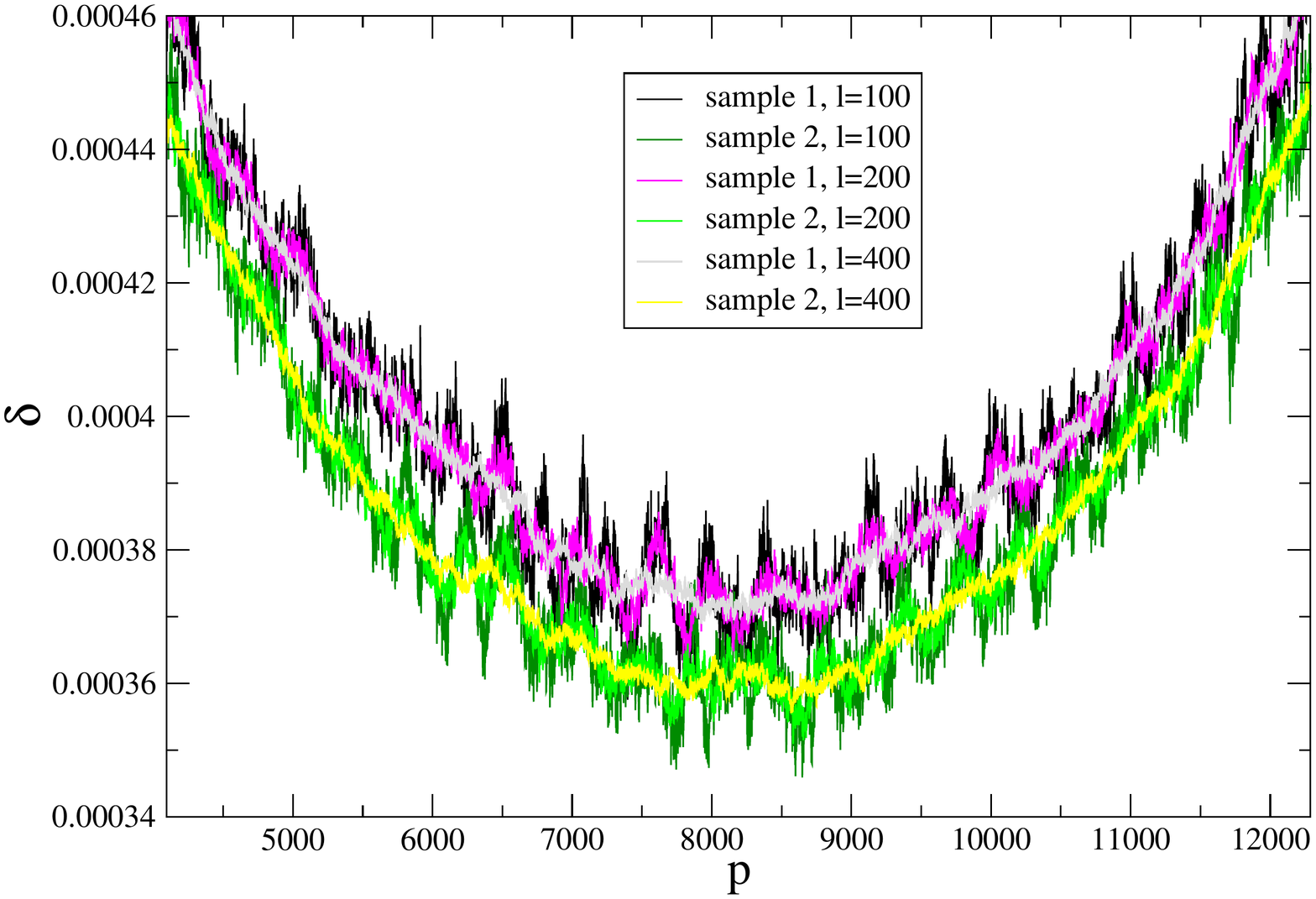}
\caption{\label{fig0a}
  The averaged level spacing $\delta_p(l) $ as function of
  $p$ the level at the center of the range $l$ over which the average is
  calculated for a {\it single} realization with different values of $l$.
  Two typical realizations of size $L=14$ and $W=2$ are presented. 
 }
\end{figure}

Even for stronger disorder where the average density of states seems smooth,
significant realization dependent structure remain. This can bee seen
in Fig. \ref{fig0a} where the averaged level spacing over a range of
eigenvalues $l$ around the $p$-th level $\delta_{l}(p)=(E_{p+l/2}-E_{p-l/2})/l$
is plotted.
Two typical realizations are presented for $W=2$, where $p$ ranges over the
the middle half of the eigenvalues, while three different values of
$l=100,200,400$ are presented. In addition to the expected
smooth global increase of $\delta$ as $p$ moves from the center of the band, 
$\delta_p(l)$ shows long range sample specific fluctuations on scales of
hundreds of levels.

\begin{figure}
\includegraphics[width=8cm,height=!]{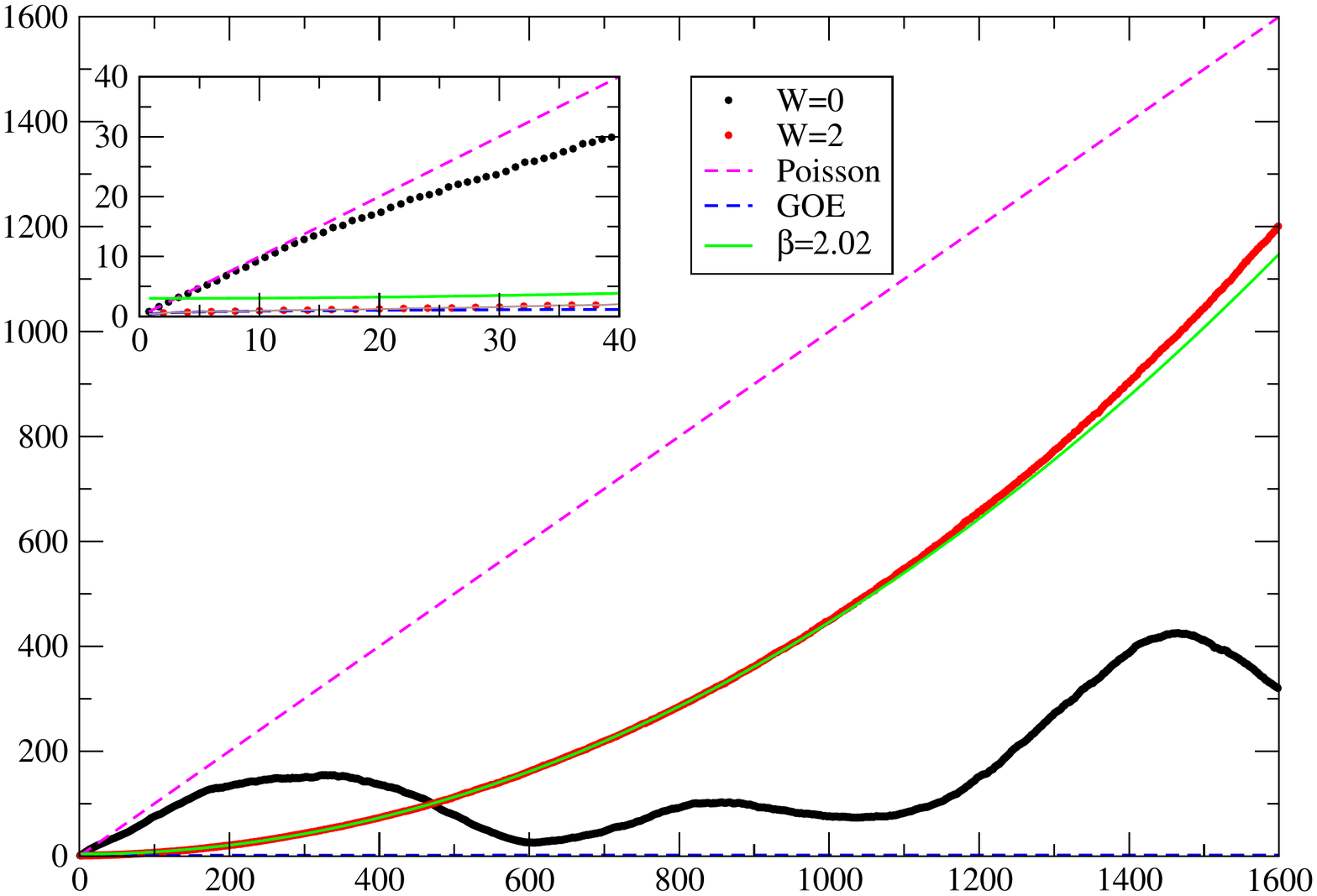}
\caption{\label{fig2a}
  The variance $\langle \delta^2 n(E) \rangle$ as function of
  $\langle n(E) \rangle$
  for $L=14$ with disorder $W=0$ and $W=2$.
  The Poisson behavior,
  $\langle \delta^2 n(E) \rangle = \langle n(E) \rangle$, and
  Wigner behavior,
  $\langle \delta^2 n(E) \rangle = (2/\pi^2)\ln(\langle n(E) \rangle)+0.44$,
are indicated by dashed curves. The inset zooms into
the small $n$ region, where the expected Poisson (for $W=0$) and Wigner ($W=2$)
behavior is seen. For larger energy scales depicted in the main figure, a completely different behavior is seen. For $W=0$ a very non-monotonous behavior
is  observed,  while for $W=2$, $\langle \delta^2 n(E) \rangle \sim
\langle n(E) \rangle^\beta$ with $\beta=2.02$ fits reasonably well.
 }
\end{figure}

Such a behavior hints towards the existence of a large scale structure of the
energy spectrum and sample to sample fluctuations.
This poses a challenge since when one studies the number
variance one would like to filter out global or sample dependence
regular behavior.
This can be problematic
since one has to separate global behavior from sample to sample fluctuations.
Let us start by a naive application of the local unfolding.
%As a first step in studying the large scale behavior of the number variance
%with local unfolding.
In order to calculate $\langle \delta^2 n(E) \rangle$, we unfold the spectrum 
by $\varepsilon_i=\varepsilon_{i-1}+2m(E_i-E_{i-1})/\langle E_{i+m}-E_{i-m}\rangle$
where $m=6$ (other values
were used with no significant change). We place the window at the
center of the band then the averages $\langle n(E) \rangle$
and $\langle n^2(E) \rangle$ are calculated over all $M$ realizations.
The results are shown in  Fig. \ref{fig2a}

The variance $\langle \delta^2 n(E) \rangle$ as function of  $\langle n(E) \rangle$
is depicted in Fig. \ref{fig2a} for $L=14$ (matrix linear size $2^{14}=16384$)
and two values of disorder $W=0$ and $W=2$. As we have seen from the
ratio statistics $r_s$ (Fig. \ref{fig2}), $W=0$ follows Poisson statistics
for small energy scales while $W=2$ follows Wigner at these scales.
Indeed, as can be seen in the inset of Fig. \ref{fig2a}, for small
values of  $\langle n \rangle$ the expected behavior of the number variance is
followed, i.e., $\langle \delta^2 n(E) \rangle = \langle n \rangle$ for Poisson
and $\langle \delta^2 n(E) \rangle = (2/\pi^2)\ln(\langle n(E) \rangle)+0.44$
for Wigner (GOE).
Nevertheless, as larger energies are examined, strong deviations
from the Poisson or Wigner behavior are observed. 

%>>>>>>>>>>>>>>>>>>>>>>>>>>>>>>>>>>>>>>>>>>>>>>>>>>>>>

The large scale behavior is very different between these two values of disorder.
For $W=0$ the linear behavior quickly  saturates, but a very non-monotonous
behavior is apparent. One cannot escape the feeling that a large scale structure
with strong sample to sample fluctuation that lurks in the spectra
is not correctly addressed by the local
unfolding. For $W=2$ the large scale behavior
is quite monotonous, shows a strong super Poissonian behavior where the
the number variance shows a power law dependence on the average number  
of states,  $\langle \delta^2 n(E) \rangle \sim \langle n \rangle^\beta$
with $\beta =2.02 \gg 1$. Although the fit seems rather decent, one must wonder
how reliable is it and whether we are seeing an artifact of the local unfolding.

\begin{figure}
\includegraphics[width=8cm,height=!]{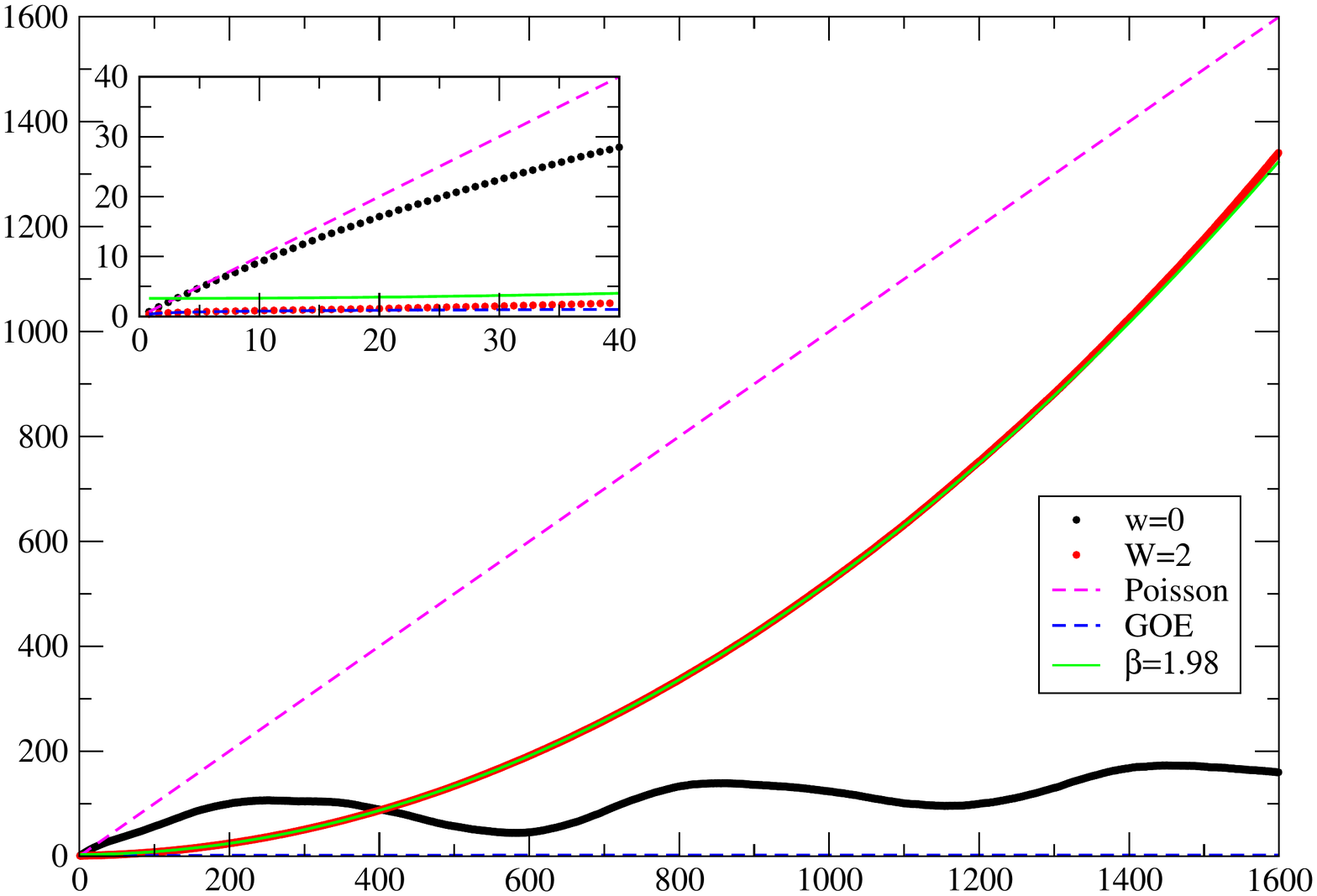}
\caption{\label{fig2b}
  As for Fig.\ref{fig2a}, with an additional  average over different positions
  of the center of the energy window. The averaging does not change much.
  For the large energy scales in the localized case ($W=0$), the non-monotonous
  behavior is somewhat dampened,  while for $W=2$, $\beta=1.98$ is similar
  to the previous result.
 }
\end{figure}

A possible cure to the sample to sample fluctuations is averaging  also
over the center of the energy window. In Fig. \ref{fig2b}
the number variance is also averaged over $21$ positions of the
center of the energy window, $\tilde E$,
equally spaced around the band center, where the
furthest point is no more than $1/15$ of the bandwidth from the center.
The number of states, $n(E,\tilde E)$, in a window of width $E$ centered
at $\tilde E$, 
is calculated, then the averages $\langle n(E) \rangle$
and $\langle n^2(E) \rangle$ are taken over all positions of the center
$\tilde  E$ and all $M$ realizations. As can be seen, the large scale
non-monotonous behavior for $W=0$ is somewhat dampened, while the behavior
for $W=2$ remains essentially the same. Nevertheless, the question remains
how much of these results are an artifact of the unfolding and sample to sample
fluctuations. We shall address global methods of unfolding  in  the next
sections.

%>>>>>>>>>>>>>>>>>>>>>>>>>>>>>>>>>>>>>>>>>>>>>>>>>>>>>

%=============================================================================

\section{Singular Value Decomposition Scree Plot}
\label{s5}

As result of these difficulties with the local  unfolding,
we change tack and use a different method to
study the spectrum, i.e., the SVD  method.
In this method no local  
unfolding is performed, and is replaced by global unfolding.
Essentially the spectrum of $M$ realizations of disorder each with $P$
eigenvalues is arranged as
a matrix $X$ of size $M \times P$ where $X_{mp}$ is the $p$ level of
the $m$-th realization. As detailed in the appendix
after carrying out SVD on 
$X$, we can write the matrix as a sum of amplitudes,
$\sigma_k$, multiplied by matrices, $X^{(k)}$, i.e., $X=\sum_k \sigma_k X^{(k)}$.
One may rank the amplitudes from the largest
to the smallest, and thus the lower values of $k$ represent modes with higher
contributions to reconstructing the  matrix. Moreover, the lower modes
tend to code the global  behavior of the matrix.  Plotting
the singular values squared $\lambda_k=\sigma_k^2$ according to their rank
is knows as the singular value scree  plot \cite{svd,svd1,svd2} and
much information can be gleaned from it. This approach has been applied to
the
spectrum of disordered systems in several studies
\cite{fossion13,torresv17,torresv18,berkovits20,berkovits21},
the first few $\lambda_k$
($k \leq O(1)$) correspond to global features of the spectra.
Higher SV ($\lambda_k$) show a power law behavior $k^{-\alpha}$.
In
the Poisson regime $\alpha=2$ for high modes, while 
$\alpha=1$ in the Wigner regime.

For the GRP model \cite{berkovits20}, the same behavior was
seen for weakly disordered (extended) and strongly disordered (localized)
regime. For the intermediate disorder NEE regime,
one expects small energy scales to show Wigner properties. Indeed,
large $k$'s follow $\alpha \sim 1$. Intermediate values of $k$, corresponding
to intermediate energy scales show unconventional behavior. They follow
a power law, but with $\alpha>2$. This super Poissonian behavior was interpreted
as the signature of the NEE phase. At small $k$
corresponding to larger energy  scales (small times) a return to an exponent of
$\alpha=2$  is observed.

%This Jives well with the picture of slow  spread
%of a wave packet for short times.

\begin{figure}
\includegraphics[width=8cm,height=!]{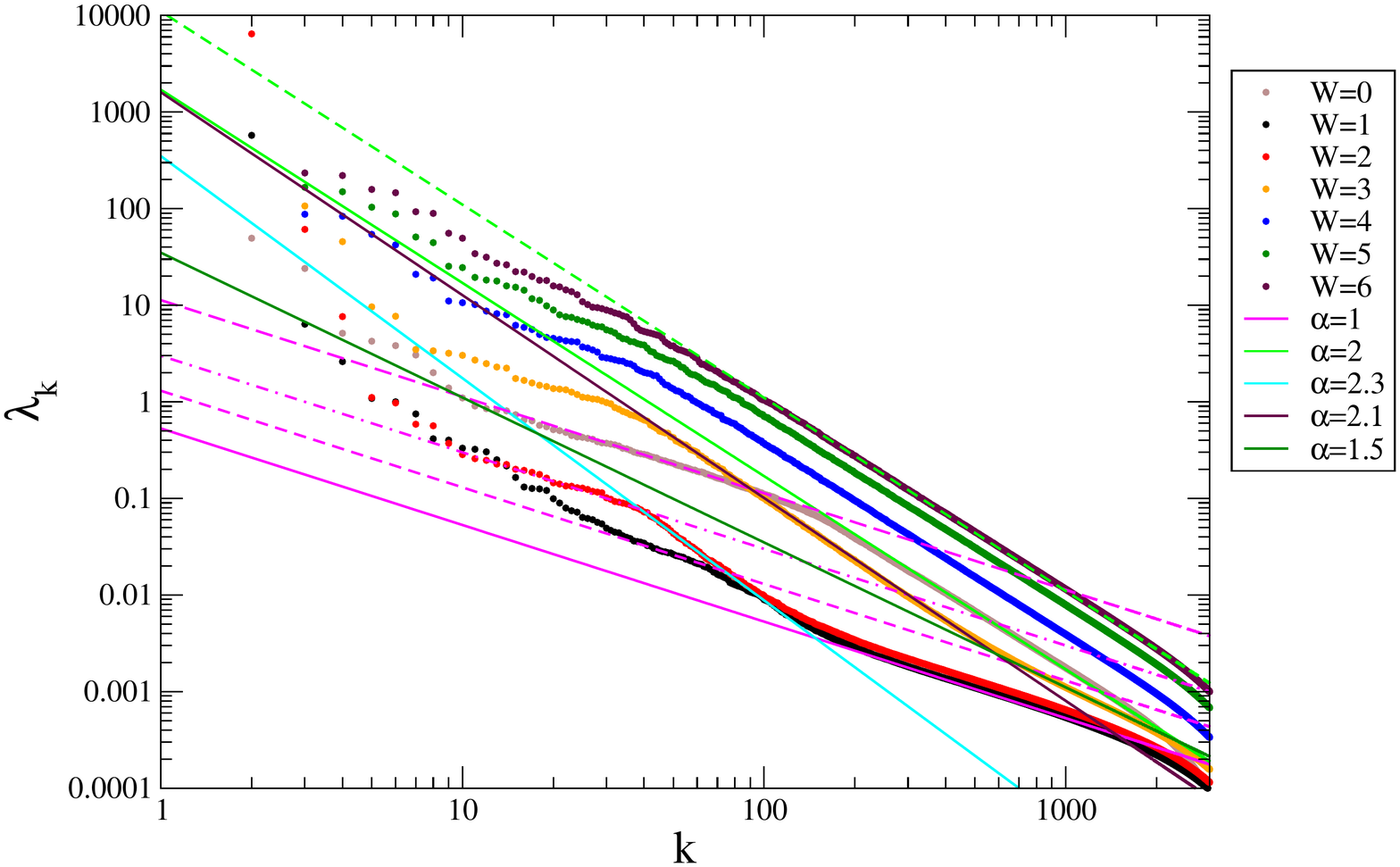}
\caption{\label{fig3}
  The scree plot of the  singular value amplitudes squared $\lambda_k$,
  where $k$ is the rank of the amplitude from highest to lowest, 
  for $L=14$ and different values of disorder $W$. $M=L^2/2=8192$ eigenvalues
  were taken from the middle of the band for $P=3000$ realizations.
  Fits to power
  laws $\lambda_k \sim k^{-\alpha}$, are depicted by the lines, where
  the magenta lines correspond to $\alpha=1$,the green lines to $\alpha=2$,
  cyan to $\alpha=2.3$,  maroon to $\alpha=2.1$  and dark green to
  $\alpha=1.5$. 
 }
\end{figure}

A somewhat similar picture emerges
for the Imbrie model. Increasing the disorder results in a change of the
dependence of the SV amplitudes on the mode number $k$.
For $W=0$ the high
$k$ values follow a power law $\lambda_k \sim k^{-\alpha}$ with $\alpha=2$, as
expected  from a localized system, matching with ratio statistics results
(Fig.\ref{fig3}). A sudden switch in the exponent to $\alpha=1$
occurs at  $k\sim  200$.
This exponent is equal to the exponent exhibited by Wigner statistics.

As can
be seen in Fig.\ref{fig3}, for weak disorder ($W=1,2$) the behavior of
the SV is quite different.
For $k>200$ and $W=1,2$ an exponent of $\alpha=1$
is evident, as  expected from systems in the Wigner (extended) regime.
This changes to an exponent larger than two
(for $W=1$, $\alpha=2.3$
for $95<k<150$; for $W=2$, $\alpha=2.3$
for $40<k<105$) for intermediate values of $k$. 
Then, similarly to $W=0$, the exponent switches back to $\alpha=1$.
Thus, in the regime
of extended behavior, the SV  amplitudes have three distinct behaviors for
different ranges of $k$. Wigner for large values  of $k$ (small energy scales,
long times), super-Poissonian ($\alpha>2$)  for an intermediate range of $k$,  
and back to $\alpha=1$.
This indicates that the extended regime in the
Imbrie model is far from trivial and signatures of different physics show up
at intermediate energy scales. This behavior is somewhat similar to
the behavior seen for the SV in the GRP model \cite{berkovits20}.
Both the GRP and Imbrie models
in the extended
regime exhibit Wigner behavior at large times
(small energy scales, large $k$), super Poissonian behavior associated with
non ergodicity at intermediate times and energy scales. For short times
(large energy scales, small $k$) the
Imbrie and the GRP models show a different behavior expressed by
different exponents ($\alpha=2$ for  GRP, $\alpha=1$ for Imbrie).
That is the result of the large scale structure
of the density
of states seen in Fig. \ref{fig0}, very clearly  for $W=0$, but still  hinted
for somewhat stronger disorder.
We shall elaborate on it in the following  section.
Thus, SVD provides support for the existence of a NEE regime for the Imbrie 
model in the weakly disordered extended regime.

For $W=3$ a crossover behavior is seen. For $700<k<2000$, $\alpha=1.5$ while
for $50<k<700$, $\alpha=2.1$, and then for $10<k<30$, $\alpha=1$. Clearly, even
for large  $k$ we do not see a clear GOE behavior expected on the basis of
the ratio statistics behavior (Fig. \ref{fig2}). The SV scree plot  behavior
seems as a crossover between Poisson and Wigner. Thus, although
finite size behavior of nearest neighbor ratio statistics unequivocally puts
the $W=3$ disorder in the Wigner regime, the larger energy scales do not
show  it. This indicates that the larger energy scales which
correspond to short times are crossing over to a closer to Poisson
behavior earlier than the short energy scales. The Wigner regime transits to
higher $\alpha$ values, while the NEE regime moves towards smaller values
of $\alpha$ closer to two. Indeed,
The $W=4$ shows an almost pure Poisson behavior although the disorder is
smaller than the critical disorder associated  with the ratio statistics.
A similar difference between the ratio statistics
corresponding to  level spacing scale and scree plot behavior
was very recently noted in Ref. \onlinecite{rao21} for the Heisenberg
chain.

In the localized regime ($W=5,6$) the expected Poisson exponent, $\alpha=2$,
is  seen for  $k>100$. For smaller values of $k$ the exponent tappers, and it
is hard to determine whether $\lambda_k$ even follows a power law at all.
Nevertheless, it looks that this region becomes smaller as $W$ increases.

\begin{figure}
\includegraphics[width=8cm,height=!]{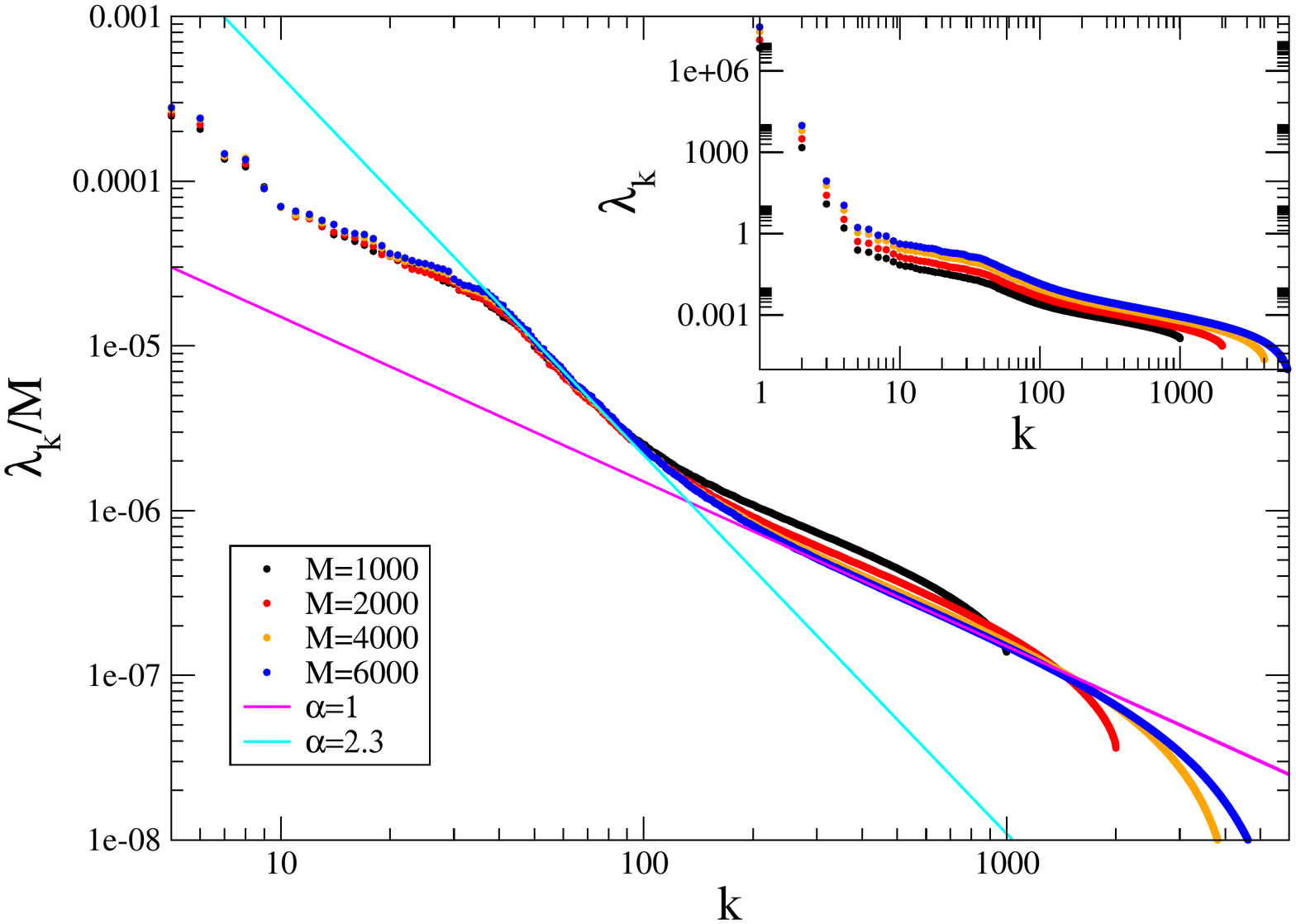}
\caption{\label{fig3.0}
  The scree plot of $\lambda_k$,
  as function of $k$ for $L=14$ and $W=2$ with $P=2^L/2=8192$ eigenvalues around the center of the band. Different number of realizations,
  $M=1000,2000,4000,6000$, are
  presented in the inset.
  In the main plot we scale the different SV amplitudes by
  dividing $\lambda_k$ by $M$. 
  Fits to a power
  laws, $\lambda_k \sim k^{-\alpha}$, are depicted by the lines, where
  the magenta line correspond to $\alpha=1$,
  and the cyan  to $\alpha=2.3$. 
 }
\end{figure}

\begin{figure}
\includegraphics[width=8cm,height=!]{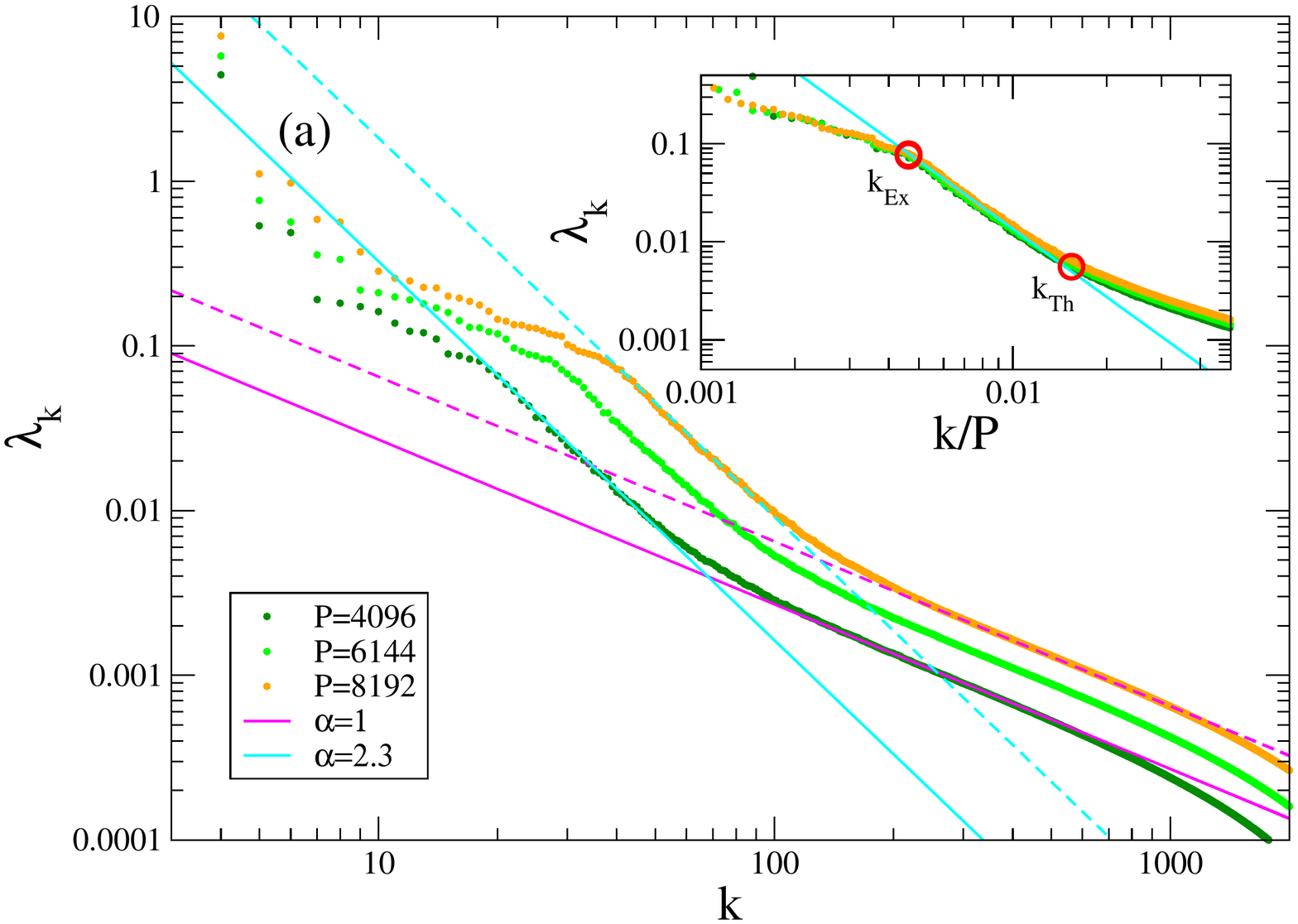}
\includegraphics[width=8cm,height=!]{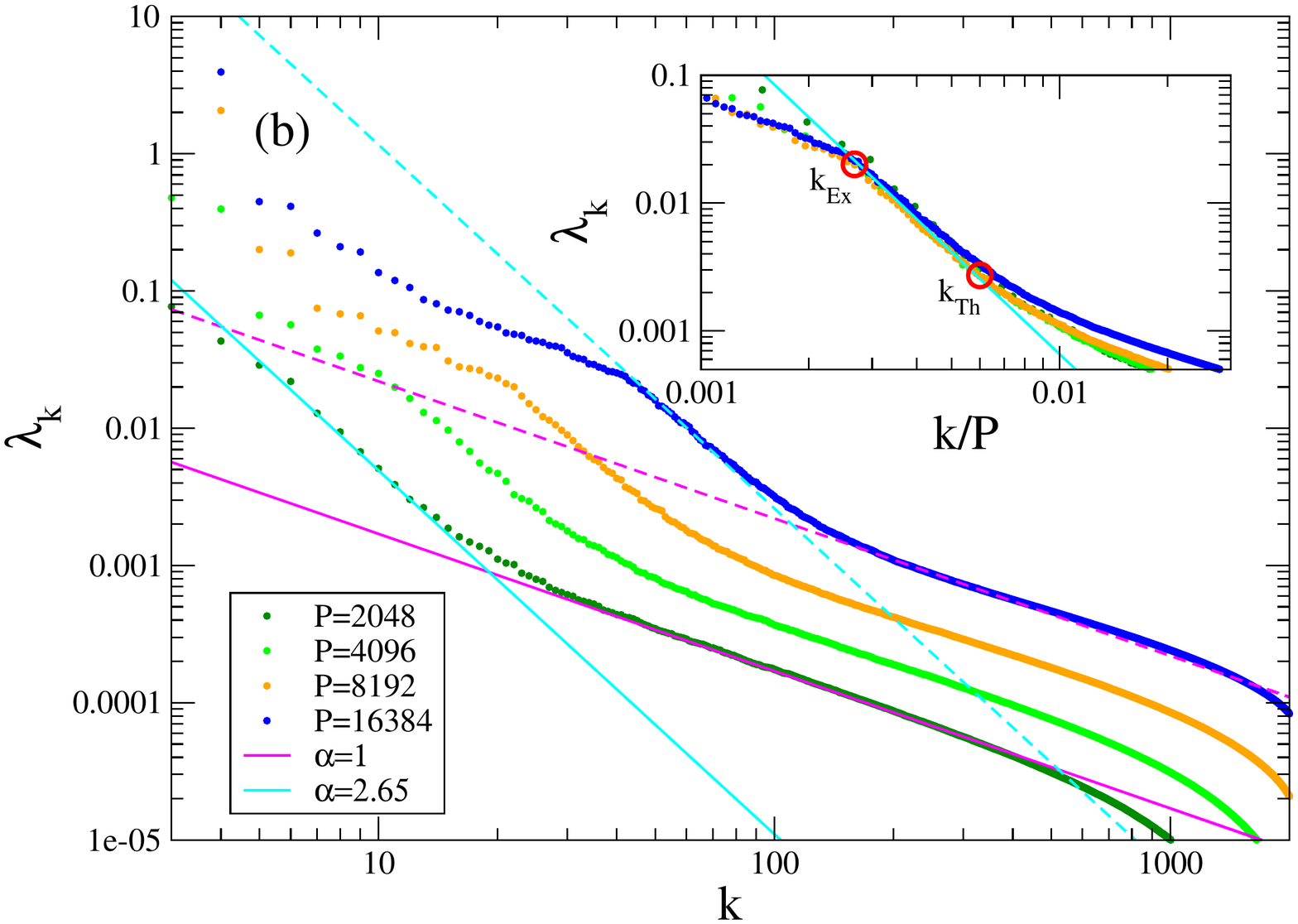}
\caption{\label{fig3.1}
  The scree plot of $\lambda_k$,
  as function of $k$ for (a) $L=14$ ($M=4096$); (b) $L=15$ ($M=2048$),
  in the weak disorder $W=2$ regime.
  Different ranges of eigenvalues $P=4096,6144,8192$  around the center
  of the band for $L=14$ and $P=2048,4096,8192,16384$ for $L=15$ are drawn.
  Fits to power
  laws are depicted by the lines, where
  the magenta lines correspond to $\alpha=1$,
  and the cyan to $\alpha=2.3$ for $L=14$,  while $\alpha=2.65$ For $L=15$.
  Inset: zoom into the region intermediate energy scale. Once the SV amplitudes,
  $\lambda_k$, are scaled by $1/P$, all curves coincide. The red circles
  indicate the position of the crossover from GOE to super Poissonian behavior
($k_{Th}$) and the
transition from the super Poissonian regime back to a Wigner regime 
($k_{Ex}$).}
\end{figure}

Returning to the weakly disordered metallic regime,
we would like to examine more carefully the
intermediate energy scale for which the super Poissonian behavior is observed.
The first issue to address is the dependence of the scree plot on the number
of realizations $M$. In Fig. \ref{fig3.0}, 
$\lambda_k$ as function of $k$ for $L=14$ and $W=2$ with
a range of $P=2^L/2=8192$ eigenvalues around the center of the band
are shown.
Four different numbers of realizations $M=1000,2000,4000,6000$
have been calculated and are presented in the inset of Fig. \ref{fig3.0}.
As the number of SV modes 
$r=\min(M,P)$ (see appendix), and here $P>M$ in all cases,
the number of modes is $r=M$.
It is clear that for small $k$'s the curves are very similar.
Rescalling $\lambda_k$ to  $\lambda_k/M$ results in all the curves falling
on top of each other for $k<100$ (Fig. \ref{fig3.0}). In the intermediate
regime $40<k<100$ for which the super Poissonian regime with an exponent
of $\alpha=2.3$ is observed the scaled curves coalesce almost perfectly.
For higher modes ($k>100$) although the curves do not coalesce (which is natural
since they terminate at different values of $k=M$), nevertheless,
the exponents are all  $\alpha=1$ for a significant range of $k$.
Thus, for a reasonable number of realizations one gets a decent
representation of large and intermediate scale behavior of the energy spectrum.

When one increases the range of eigenvalues, $P$, while keeping the number
of realization $M$ fixed it is possible track the two energies
determining the crossover from GOE to super Poissonian behavior, $E_{Th}$,
and the
transition from the super Poissonian regime to a Poissonian regime,
$E_{Ex}$. One might expect that since SVD modes describe the energy
spectrum of width $P \delta$,
resulting in the $k$th mode corresponding to a $P \delta/k$ energy range.
Thus the position  of $k_{Th}$, the mode for which the exponent changes
should depend linearly on $P$. 
Indeed, from Fig. \ref{fig3.1} which presents a scree plot of the SV
of $L=14$  ($2^L=16384$) and $L=15$ ($2^L=32768$) deep in
the weak disorder regime ($W=2$) for
$M=4096$ ($L=14$) or $M=2048$ ($L=15$) realizations, and different
ranges of eigenvalues
$P$ centered around the middle of
the band, one can see that the 
SV amplitudes, $\lambda_k$, scale as $1/P$. As can be seen in the
insets,  all curves coincide after rescaling. Estimating the energy scales
from the scree plots leads to:
$E_{Th}  \sim P \delta/k_{Th} \sim 80 \delta$ ($L=14$)
and $E_{Th}  \sim 160 \delta$ ($L=15$).
Similarly, $E_{Ex} \sim M \delta/k_{Ex} \sim 200 \delta$ ($L=14$) and
$E_{Ex} \sim P \delta/k_{Ex} \sim 400 \delta$ ($L=15$).
Since,
%$k_{Th}$ and $k_{Ex}$ scale with $P$
roughly speaking, $\delta \sim B/2^L$ , (where $B$ is the band
width which depends only  weakly on $L$), one may postulate that
$E_{Th}$  and $E_{Ex}$ correspond to a fixed fraction of the band width
for the same disorder.
The values of
$E_{Th}$ and $E_{Ex} $ are within the same ranges for which we observed
the large scale structure in Fig. \ref{fig0a}.

\begin{figure}
\includegraphics[width=8cm,height=!]{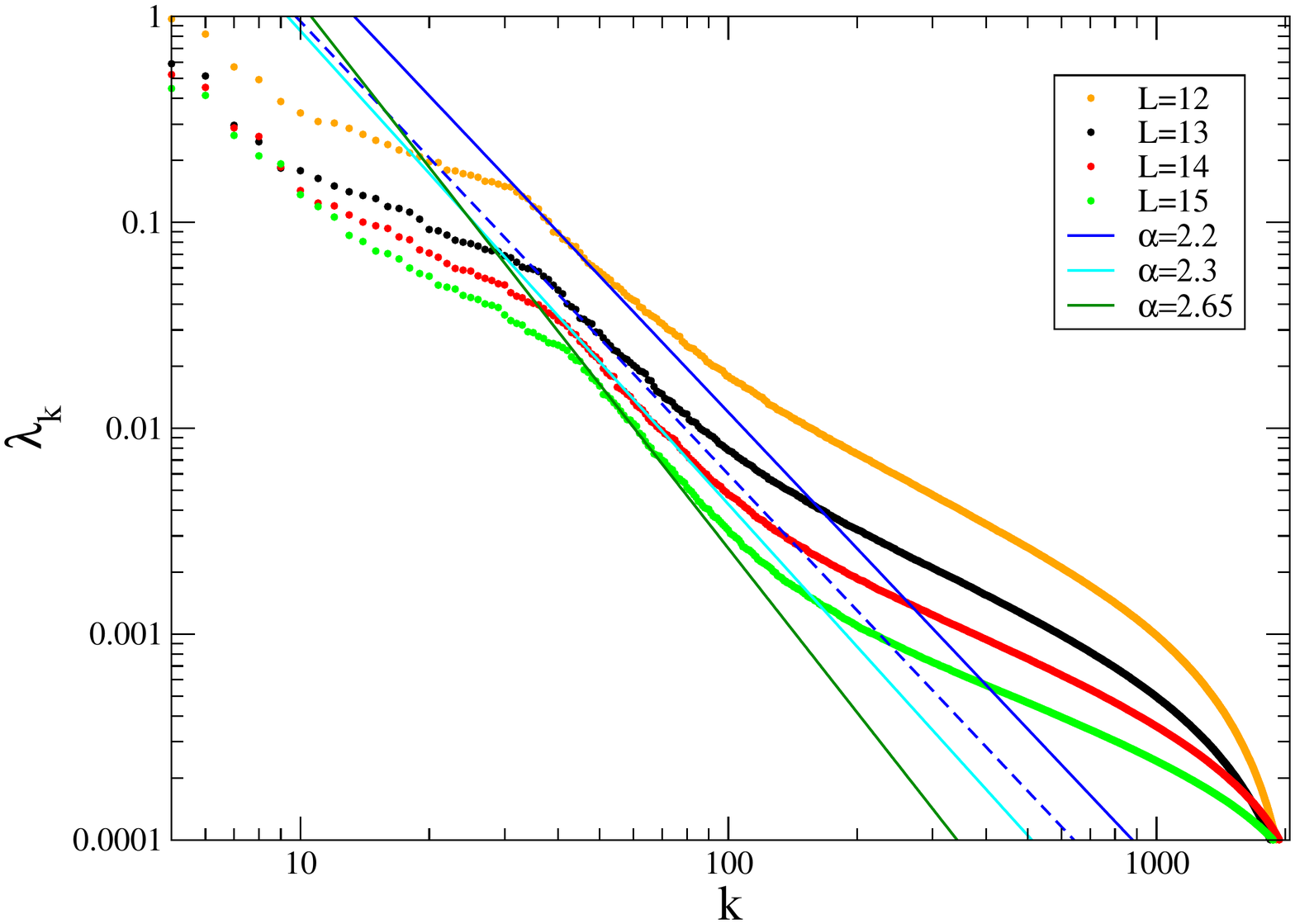}
\caption{\label{fig4}
  The scree plot of $\lambda_k$,
  as function of $k$ for $W=2$ and four sizes
  of $L=12,13,14,15$ with $M=2048$
  realizations. In all cases $P=2^L/2$.
  Fits to power
  laws $\lambda_k \sim k^{-\alpha}$, are depicted by the lines, where
  the blue line correspond to $\alpha=2.2$ (fit $L=12,13$),
  the cyan  to $\alpha=2.3$ ($L=14$), and the dark green to $\alpha=2.65$
  ($L=15$). The transition between the universal and super Poissonian
  behavior, $k_{Th}$, and between the super Poissonian and large scale behavior,
  $k_{Ex}$, is similar for all sizes. 
 }
\end{figure}

In Fig. \ref{fig4} we probe the influence of size, $L$,
on the intermediate region. Here $L=12,13,14,15$, $W=2$ and $M=2048$
realizations are considered for all sizes. In all cases $P=2^L/2$  (half of the
eigenvalues around the middle of the band). The exponent in the intermediate
energy range increases as the size becomes larger. For $L=12,13$ the exponent
$\alpha=2.2$, for $L=14$ its $\alpha=2.3$, and for $L=15$, the largest size
considered here, $\alpha=2.65$. One may conclude that the intermediate
super Poissonian
behavior is enhanced  by the increase of the system size. Moreover,
the crossover regions between the regions becomes sharper and $k_{Th}$ and
$k_{Ex}$ easier to pinpoint as $L$ increase. It can be also seen that
for all  sizes the Thouless and large scale energy scales do not vary much,
in line with our previous conclusion that they depend on the band width.

Although $E_{Th}$ moves to lower energies as $W$ increases (at least when $W$
approaches the transition value) similar to the Thouless energy for the
single particle Anderson model there are nevertheless important differences
for the larger energy scales. Indeed,
for both Anderson localization and MBL (or GRP) the behavior on larger
scales is super Poissonian ($\alpha>2$), but there are two main differences.
The first is that while for the Anderson case $\alpha$ depends mainly on
dimension and only weakly on $W$ and not at all on $L$, for the MBL
model the power law has a very strong dependence on  disorder and system size.
The second difference is that for the MBL case an additional energy scale
($E_{Ex}$) is evident, while for the Anderson model it is absent.

Thus, the super Poissonian regime seems robust and not a fluke of the range
of eigenvalues considered or small size. Nevertheless, from the available
data it is not possible to  extrapolate what is the $\alpha$  value at infinite
size.

\section{Singular Value Decomposition Global Unfolding}
\label{s6}

Another way that SVD can be used, is to apply its results for unfolding the
spectra and then perform a standard number variance calculation. The unfolding
is based on reconstructing the matrix $X$ where the first (or few)
contributions  of  the SV decomposition are dropped since they encode
the global behavior. Specifically, capturing the global behavior of  the
energy spectrum by
$\tilde X_{lp} =\sum_{k=m}^r \sigma_k X_{lp}^{(k)}$
(see appendix) with $m$ determined by examining the
scree plot and identifying the point  where the first few modes change
the behavior. For example in the scree plot for the $L=14$, $W=2$  case
seen in the inset  of Fig. \ref{fig3}, one chooses $m=4$.

\begin{figure}
\includegraphics[width=8cm,height=!]{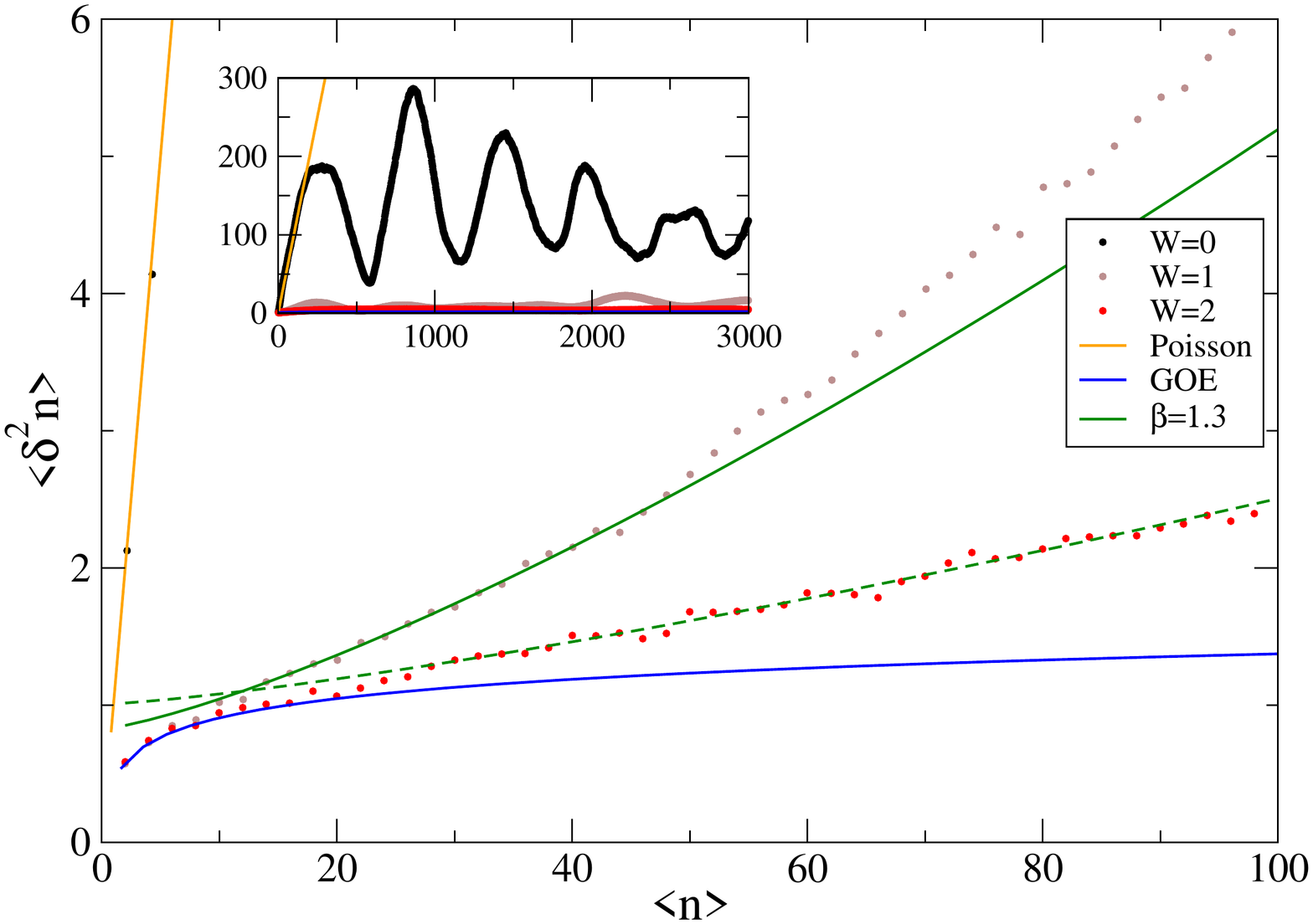}
\caption{\label{fig5}
  The number  variance 
$\langle \delta^2 n(E) \rangle$ as function of
  $\langle n(E) \rangle$ for weak disorder ($L=14$, $W=0,1,2$,
  $P=4096$ and $M=4096$).
  The Poisson and Wigner behaviors correspond to the orange and blue
  curves correspondingly. The inset presents the whole range of
  $\langle n(E) \rangle$
  values, while the main figure zooms into smaller $\langle n(E) \rangle$.
  The green curves correspond to $\langle \delta^2 n(E) \rangle \sim
  \langle n(E) \rangle^\beta$ with $\beta=1.3$ and the full and  dashed line
  represent different prefactors.
%  (b) Strong disorder ($W=3,4,5,6$).
%  The green curves correspond to fits of $\langle \delta^2 n(E) \rangle \sim
%  \langle n(E) \rangle^\beta$ with $\beta \sim 1$.
 }
\end{figure}

Defining the global unfolded $l$-th eigenvalue of the $p$-th realization as:
\begin{eqnarray} \label{global}
\tilde \varepsilon_l=
\tilde \varepsilon_{l-1}+\frac{X_{lp}-X_{l(p-1)}}{\tilde X_{lp}-\tilde X_{l(p-1)}}+1,
\end{eqnarray}
and calculating the number variance centered on the middle of the unfolded
spectra, results in the number variance presented in Fig. \ref{fig5}.
Here we focus on the weak disorder regime.
First, lets examine the behavior of the number variance
for small average numbers, corresponding to small energy scales
shown  in the main panel.
For $W=0$ we see a close to linear behavior
with $\langle \delta^2 n(E) \rangle = \langle n \rangle$.
For $W=1,2$ we see see in the inset a Wigner (GOE) behavior,
$\langle \delta^2 n(E) \rangle = (2/\pi^2)\ln(\langle n(E) \rangle)+0.44$,
which holds up to $\langle n(E) \rangle \sim 10$ for $W=1$ and
$\langle n(E) \rangle \sim 20$ for $W=2$.

For large energy scales, a different dependence emerges.
The variance saturates with quasi-periodic oscillations which are very
pronounced for small  disorder and dampened at higher $W$.
This  behavior conforms to 
the low modes (small $k$) power law seen for $\lambda_k$
(Fig. \ref{fig3}). In this region $\alpha=1$, and assuming
$\langle \delta^2 n(E) \rangle \sim \langle n(E) \rangle^{\beta=\alpha-1}$, 
\cite{berkovits21},
leading to the expectation that the  the power law
behavior of the number variance for large energy scale will correspond to
$\beta=0$, i.e., saturation. Nevertheless, on top of the saturation
a quasi-periodic oscillations is observed. This is the
result of the finite range of $k$ for which the exponent is equal to $1$. 

For the weakly disordered regime ($W=1,2$) the long time (small  energy scales)
Wigner behavior is followed by an intermediate time and energy scale for which
$\langle \delta^2 n(E) \rangle \sim \langle n(E) \rangle^{\beta}$, and
$\beta=1.3$ (for $W=1$ fit to  the range $10<\langle n(E) \rangle<50$, and
for $W=2$ to $30<\langle n(E) \rangle<100$). This corresponds to 
powers larger than $2$ we have seen in the scree plot for the SV amplitudes.
Thus, this regime  corresponds to $E_{Th}<\langle n(E) \rangle \delta <E_{Ex}$.
The estimation in the previous section (assuming a factor two)
for $L=14$ and
$W=1$  of $E_{Th}  \sim 20 \delta$ and  $E_{Ex} \sim 40 \delta$,
while for $W=2$, $E_{Th}  \sim 40 \delta$ and  $E_{Ex} \sim 100 \delta$.
These estimations fits reasonably well 
the range of the super Poissonian behavior seen for the globally unfolded
number variation.
Moreover, $\beta=\alpha-1=1.3$ in line
with the behavior of the exponent observed for the SV in the region
of $k_{Th}< k <k_{Ex}$.

%The $W=3$ strength of disorder exhibits transitional behavior.
%While the nearest neighbor ratio statistics
%clearly shows that $W=3$ is deep in the extended Wigner regime, the
%globally unfolded number variation is more similar to the Poisson (actually
%critical) behavior, with a  exponent close to one ($\beta=1.08$).
%Nevertheless, contrary to the stronger disorder behavior, there is
%also a transitional signature in the high $k$ behavior 
%scree plot (Fig. \ref{fig3}) where for for $50<k<700$, with $\alpha=2.1$.
%corresponding to $E_{Th} \sim M \delta/ 2 k_{Th} = 6 \delta$ (incorporating
%the factor $2$ seen for the behavior for $W=2$) and
%$E_{sb}= 80 \delta$ in a reasonable agreement with the regime for whic
%the fit $\beta=1.08$ holds  and with the estimation of $\beta=\alpha-1=1.1$.
%
%For stronger disorder ($W=4,5,6$), the
%globally unfolded number variation at large energy scales
%but the the range of levels is to small to say anything definitive. Since the
%range covers a quarter of the total of levels in the band, it is impossible
%to asses the behavior without increasing the systems size.

\section{Discussion}
\label{s7}

In the previous sections it has been shown that
the energy spectra of the quantum random antiferromagnetic Ising chain
with mixed transverse and longitudinal fields displays a clear signature of a
super Poissonian behavior for a range of energies $E_{Th}<E<E_{Ex} $.
The super Poissonian behavior appears deep in  the metallic regime and its
range grows as the system approaches the MBL transition. On the other hand,
the characteristics of the region approach a regular Poisson behavior
(a power of two in the scree  plot) as one gets closer to  the localized
regime. The super Poissonian behavior becomes more pronounced as the system
size increases, and exhibits scaling behavior as function of the number
of realizations and range of eigenvalues considered. Thus, the super Poissonain
regime is robust and does not seem to be an artifact of small systems, although
it is hard to extrapolate to much larger systems. This
behavior is brought to light once global unfolding and sample to sample
fluctuations are taken into account using the SVD method, both by scrutinizing 
the scree plot of the SV amplitudes as well as studying the number variance of
the spectra after unfolding the spectra by SVD.

Thus, for a model which is one of the canonical microscopical
model for studying the MBL transition, the metallic
phase is far from trivial. The small energy scales show all the universal
features expected in the metallic regime, while higher energy scales clearly
are non-universal.
A non Poissonian behavior at large energies
has also been very recently seen for an other
canonical microscopic model for MBL, the
Heisenberg chain \cite{rao21}. Nevertheless, $E_{Ex}$ is not
observed there since only a 
small range of eigenvalues were considered there, similar to the $P=2048$
case depicted in Fig. \ref{fig3.1}b. 
As in itself the deviation from universal behavior of the spectrum at
larger energy scales is not surprising, since a somewhat similar
deviation from the universal behavior of the energy spectra is seen 
in the single particle energy spectrum and associated with the Thouless energy.
There the reason for the termination of the universal behavior is very clear.
At short times (corresponding to large energy scales) diffusive behavior has
not had time to evolved and experience the whole sample
and therefore the behavior is not yet universal. 
For the MBL model the crossover from the short time behavior does
not occur directly to the diffusive (universal) behavior, but there is
an intermediate times for which the motion of a wave packet is extended,
but nevertheless it does not cover the whole
phase space and only on longer times it crosses over to the diffusive
regime. Both crossovers leave a distinct signature  in the energy spectrum
and establishes energy scales ($E_{Ex}$ and $E_{Th}$) which can be extracted
using SVD. This regime exists only in the metallic regime, while in the
localized regime there is only a transition from non-universal short times 
behavior to a localized behavior.

The origin of this intermediate energy (or time) regime is not clarified
by this study. Whether is stems from the structure of the
coupling of states in the
Fock space resulting in a quantum random graph, or other explanations which
hinge on static or dynamical rare regions
in the system such as Griffiths regions which may drive KT transitions, needs
more study. Of course clarifying the finite size scaling of the
intermediate regime is highly desirable, but unfortunately seems beyond
current and reasonable future numerical capabilities.
A possible continuation to this study would be the study of the
energy spectrum of other models models with a different geometry than the
1D chains, such as the a random network or a modified SYK model. Although
one will continue to suffer from the constrains of small systems, one will have
freedom of tweaking geometry which may help understanding the physics behind
this intermediate region.

%\begin{acknowledgments}
%[Useful discussions with A. Turner and G. Murthy are gratefully acknowledged.
%\end{acknowledgments}

\appendix*
\section{Singular Value Decomposition}

The singular value
decomposition (SVD) \cite{svd,svd1,svd2} is a  method to decompose a matrix
$X$ of size $M \times P$ ($X$  is not necessarily Hermitian nor square)
into a sum of matrices. 
The matrix $X$ represents data arranged by rows and columns, where the
arrangement depends on the application.
For the SVD analysis of the spectrum
one writes the $M$ realizations of disorder and the $P$ eigenvalues each,
as a matrix $X$ of size $M \times P$ where $X_{mp}$ is the $p$ level of
the $m$-th realization. The matrix
$X$ is decomposed to $X=U \Sigma V^T$, where
$U$ and $V$  are $M\times M$ and $P \times P$ matrices correspondingly,
and $\Sigma$ is a {\it diagonal} matrix of size $M \times P$ and rank
$r=\min(M,P)$. The $r$ diagonal elements of $\Sigma$, denoted as $\sigma_k$
are the singular values (SV) of the matrix which are positive
and could be ordered by their size such that
$\sigma_1 \geq \sigma_2 \geq \ldots \sigma_r$.
The Hilbert-Schmidt norm of the matrix
$||X||_{HS}=\sqrt{Tr X^{\dag}X}=\sum_k \lambda_k$ (where $\lambda_k=\sigma_k^2$).
Therefore, using the SVD
the matrix $X$ could be written as a series composed of matrices $X^{(k)}$,
where $X^{(k)}_{ij}=U_{ik}V^T_{jk}$ and $X_{ij}=\sum_k \sigma_k X^{(k)}_{ij}$.
Thus, this series in an approximation
of matrix $X$, where the sum of the first $m$ modes gives a matrix
$\tilde X =\sum_{k=1}^m \sigma_k X^{(k)}$, for which $||X||_{HS}-||\tilde X||_{HS}$
is minimal.

\end{document}